\documentclass[aps,pre,twocolumn,superscriptaddress]{revtex4-1}
\usepackage{epsfig,dsfont,amssymb,amsmath,amsthm,amsfonts,amsbsy,mathrsfs}
\usepackage{graphicx}
\usepackage{color}
\usepackage{bm}
\usepackage{multirow}
\usepackage{natbib}

\usepackage{tikz}
\usetikzlibrary{arrows,shapes,chains}

\begin{document}
\title{Softness, anomalous dynamics, and fractal-like energy landscape in model cell tissues}
\author{Yan-Wei Li}
\affiliation{Division of Physics and Applied Physics, School of Physical and
Mathematical Sciences, Nanyang Technological University, Singapore 637371, Singapore}
\affiliation{School of Physics, Beijing Institute of Technology, Beijing 100081, China}
\author{Leon Loh Yeong Wei}
\affiliation{Division of Physics and Applied Physics, School of Physical and
Mathematical Sciences, Nanyang Technological University, Singapore 637371, Singapore}
\author{Matteo Paoluzzi}
\affiliation{Departament de Física de la Mat\`eria Condensada, Universitat de Barcelona, C. Martí Franqu\`es 1, 08028 Barcelona, Spain}
\author{Massimo Pica Ciamarra}
\email{massimo@ntu.edu.sg}
\affiliation{Division of Physics and Applied Physics, School of Physical and
Mathematical Sciences, Nanyang Technological University, Singapore 637371, Singapore}
\affiliation{
CNR--SPIN, Dipartimento di Scienze Fisiche,
Universit\`a di Napoli Federico II, I-80126, Napoli, Italy
}
\date{\today}


\begin{abstract}
Epithelial cell tissues have a slow relaxation dynamics resembling that of supercooled liquids. Yet, they also have distinguishing features.
These include an extended short-time sub-diffusive transient, as observed in some experiments and recent studies of model systems, and a sub-Arrhenius dependence of the relaxation time on temperature, as reported in numerical studies. 
Here we demonstrate that the anomalous glassy dynamics of epithelial tissues originates from the emergence of a fractal-like energy landscape, particles becoming virtually free to diffuse in specific phase space directions up to a small distance.
Furthermore, we clarify that the stiffness of the cells tunes this anomalous behaviour, tissues of stiff cells having conventional glassy relaxation dynamics.
\end{abstract}

\maketitle
\section{Introduction}
Cells in tissues rearrange in many biological processes, including embryonic development, wound healing, and tumour metastases~\cite{Cornelis, Silberzan_pnas, Levine_PNAS,trepat2018mesoscale}. 
The resulting tissue dynamics is slow and heterogeneous in both space and time, 
as cells in tissues may spend long transients in the cages formed by their neighbours, before relaxing through cooperative processes~\cite{Manning2013, Weitz_PNAS2011, Bi_prx}.
These observations evidence strong similarities between the dynamics of tissues~\cite{PhysRevLett.104.168104,Manning2013, Weitz_PNAS2011, Bi_prx, Manning_EPL, EMT1, EMT2, EMT3} and that of supercooled liquids~\cite{Stillinger_Nature, Binder2011}.
However, the investigation of the relaxation dynamics of model cell tissue~\cite{Manning_EPL, 2020arXiv200714107S} where temperature-like stochastic forces drive particles, revealed distinct features not shared by ordinary supercooled liquids. 
In particular, the relaxation time was observed to grow as $\tau_\alpha \propto \exp(-E/T^x)$ with 
$x < 1$, a sub-Arrhenius behavior markedly distinct from the strong, $x=1$, or a super-Arrhenius, $x>1$, behavior of supercooled liquids~\cite{angell1995formation}.
Besides, while in supercooled liquids particles do not diffuse during the transient caging-regime, the mean square displacement scaling as $t^\alpha$ with $\alpha \to 0$ at low temperature, cells in tissue may transiently exhibit a sub-diffusive behaviour, during which $1 > \alpha > 0$ is approximately constant.
Experimental findings are compatible with this sub-diffusive behavior~\cite{Rogers2007, Manning2013, Nixon-Abell2016, Armiger2018, Fodor2018}, which occurs on a short time scale where the dynamics is likely to be dominated by thermal effects, rather than by self-propulsion of cells.
These results are indicative of unusual features of the energy landscape of tissues, which have not yet been rationalized. 
It has not even been ascertained if the anomalous glassy dynamics ubiquitously occur in epithelial cell tissues, or rather if it depends on the mechanical properties of the cell, recently correlated to their geometrical features~\cite{Park_NM}.

In this paper, we show that the distinctive sub-diffusive and sub-Arrhenius glassy relaxation only occur in a tissue of highly deformable cells, while conversely a conventional glassy relaxation dynamics occurs. 
Furthermore, we rationalize that these distinct features signal the existence, in the energy landscape of highly-deformable epithelial tissues, of {\it selected} phase space directions along which the system moves almost freely, for short distances. Displacements along these phase space directions trigger cell rearrangement processes, or T1 transitions~\cite{Rivier_1984, Staple2010, Bi_softMatter}, that have a negligible energy cost.
The physical mechanism leading to sub-diffusion  establishes an unexpected connection between the dynamics of cell tissues and that of particles diffusing in random media, and indicates that the energy landscape of tissues is locally fractal-like as that of the Lorentz model close to the percolation threshold~\cite{Henk1982, Erwin2006, Zeitz2017}.

\section{Voronoi model}
\subsection{Numerical model}
We investigate the dynamics of a model of epithelial tissues~\cite{Frank_vertex,  Staple2010, Bi, Manning1,Vertex_model, Bi_prx}, where the configurational degrees of freedom are the centers of mass of the cells, $\{\mathbf{r}_i\}$, and the shape of cell $i$ is that of the Voronoi cell centered in $\mathbf{r}_i$.
Biological considerations~\cite{Frank_vertex,  Staple2010, Bi, Manning1,Vertex_model, Bi_prx,PhysRevLett.120.268105,giavazzi2018flocking} indicate that the mechanical energy of a cell depends on its area $A_i$ and perimeter $P_i$, $E_i = K_A (A_i-A_{0}^i)^{2}+K_P(P_i-P_{0}^i)^{2}$, where $A_0^i$ and $P_0^i$ are preferred values, while $K_A$ and $K_P$ are area and perimeter elastic constants.
Hence, the dimensionless energy functional is
\begin{equation}
e=\sum_{i=1}^{N}[(a_{i}-a_{0}^{i})^{2}+r^{-1}
(p_{i}-p_{0}^{i})^{2}],
\label{eq:E}
\end{equation}
where the sum runs over all $N=1024$ cells of the system, $a_i = A_i/l^2$ and $p_i = P_i/l$ with $l$ the unit of length which we have chosen so that $\langle a_{i} \rangle=1$.
The preferred area $a_{0}^{i}$ is uniformly distributed in the range $0.8$--$1.2$, to avoid crystallization, while the preferred perimeter is fixed to $p_{0}^{i}=p_0\sqrt{a_{0}^{i}}$, with $p_0$ the target shape index. 
The non-dimensional energy then depends on the inverse perimeter modulus, $r = K_A l^2/K_P$, we fix to 1, and on $p_{0}$. 
This parameter determines the cell deformability, higher values of $p_0$ corresponding to more deformable cells~\cite{Bi, Ourwork_PRM}. Simulations are performed using periodic boundary conditions.

\subsection{Connection with experiments}

The use of this model to simulate epithelial tissues poses two challenges. First, one would need to determine the values of the model parameters.
While the physical and biological interpretation of the model's parameters is clear~\cite{Frank_vertex,Staple2010}, these have never been experimentally determined. 
Estimating these parameters, and in particular the elastic constants, would require probing the interaction of cells in a tissue, and take into account that the model focuses on a two-dimensional representation.
Secondly, one needs to drive the cells via active forces.
This is a issue as the features of the active forces inducing the dynamics of cell tissues are still unclear, and indeed different models have been proposed in the literature~\cite{Barton2017, Bi_prx}. 
Specifically, the issue is whereas the biological process leading cell motion also induces aligning interactions between the cells~\cite{Barton2017}.

To tackle these issues, we perform simulations in the NVT ensemble, where $T$ should be interpreted as an effective temperature. Specifically, we integrate the equations of motion via the Verlet algorithm, and fix the temperature using a Langevin thermostat~\cite{Allen_book}.
Furthermore, we relate the model parameters to experimental values considering their effect on the dynamics.
While the diffusion coefficient of cells in epithelial tissues vary greatly with the control parameters, a typical order of magnitude estimate is
$D \simeq 10^{-2} \mu m^2/{\rm min}$ (e.g, ~\cite{Armiger2018,Dieterich2008}).
We investigate effective temperature values leading to a diffusion coefficients, which we estimate from the mean square displacement of our numerical model, of order $D_{\rm sim} = 10^{-3} l^2/\tau_0$, with $l$ and $\tau$ are length and time units. This is also an order of magnitude estimate, as the diffusion depends on the temperature.
Equating the numerical and the experimental diffusion coefficient, and fixing our length unit to the typical cell size, of order $10\mu m$, we estimate our time unit to be approximately $0.06$s. 

In the following, we investigate the dynamics for $t \leq 10^5 \tau_0 \simeq 4h$.
This is a short time scale with respect to that of biological processes such as cell reproduction and apoptosis, which affect cell size \cite{Puliafito_PNAS,Straetmans2019}, and with respect to the time scale of cell volume fluctuations~\cite{Zehnder2015}, which we therefore neglect.

\section{Conventional and anomalous glassy dynamics: $p_0$ dependence}
\begin{figure}[t]
 \centering
 \includegraphics[angle=0,width=0.5\textwidth]{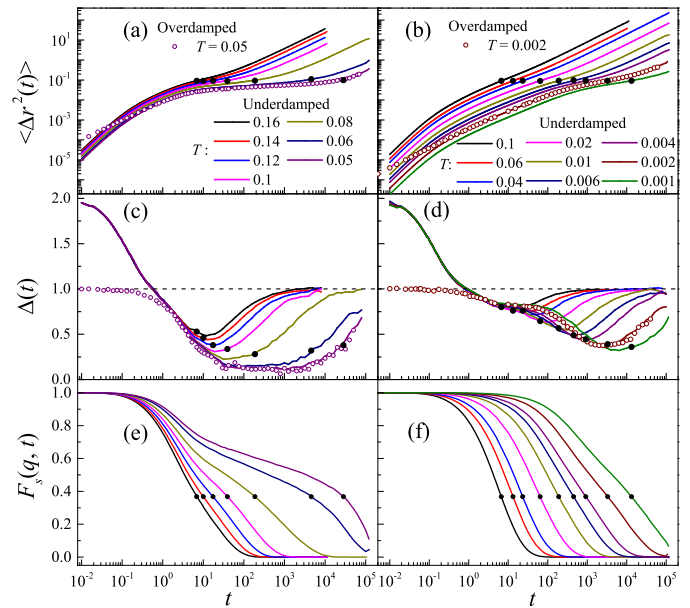}
 \caption{Time dependence of the mean square displacement (a), of its log-slope (c) and the self-intermediate scattering function (e), at $p_0=3.0$. (b), (d) and (f) show the same quantities at $p_0=3.81$. 
Open symbols in (a)-(d) are results obtained via overdamped simulations, for the indicated low-temperatures values. In all panels, black dots mark the relaxation time. 
\label{fig:dyn}
}
\end{figure}

At zero temperature, on increasing the target shape index, this model exhibits a sharp crossover for $p_0 \simeq 3.81$, which is reminiscent of a rigidity transition~\cite{Bi, Ourwork_PRM,sussman2018no}.
Here, we compare the relaxation dynamics at $p_0 = 3.0$ and at $p_0 = 3.81$, respectively in the solid phase and close to the crossover. 
We investigate the MSD $\left\langle\Delta r^{2}(t)\right\rangle=\left\langle\frac{1}{N} \sum_{i=1}^{N} \Delta \mathbf{r}_{i}(t)^{2}\right\rangle$, with $\Delta \mathbf{r}_{i}(t)$ displacement of particle $i$ at time $t$, its log-slope $\Delta(t)=\mathrm{d}\left(\ln \left\langle\Delta r^{2}(t)\right\rangle\right) / \mathrm{d}(\ln (t))$, and the self-intermediate scattering function (ISF) $F_{s}(q, t)=\left\langle\frac{1}{N} \sum_{j=1}^{N} e^{i\mathbf{q} \cdot \Delta \mathbf{r}_{j}(t)}\right\rangle$ with $q=|\mathbf{q}|$ the wavenumber of the first peak of the static structure factor. We define the relaxation time $\tau_\alpha$ as $F_{s}(q, \tau_\alpha)=e^{-1}$.

Stiff cells ($p_0=3.0$) exhibit a typical glassy behaviour~\cite{Stillinger_Nature, Binder2011};
As the temperature decreases the MSD develops an increasingly long plateau during which $\Delta(t)$ attains a small value, and the ISF develops a two-step decay (Figs.~\ref{fig:dyn}(a), \ref{fig:dyn}(c) and \ref{fig:dyn}(e)). 
Conversely, soft cells ($p_0=3.81$) relax in a qualitatively different way. 
Although the dynamics slows down dramatically at low temperatures, the MSD does not exhibit a true plateau, if not at extremely low temperature, and the ISF does not decay in two steps (Figs.~\ref{fig:dyn}(b) and \ref{fig:dyn}(f)).
More importantly, an extended sub-diffusive behaviour follows the short-time ballistic regime one. 
Indeed, the log-slope $\Delta(t)$ of the mean square displacement develops an extended plateau, we show in Fig.~\ref{fig:dyn}(d).
These results clarify that an anomalous glassy dynamics occurs for soft cells, as previously observed~\cite{Manning_EPL}, but not for stiff ones.
The stiffness of the cells, therefore, does not simply alters the energy scale for particle rearrangement, but rather qualitatively influences the relaxation dynamics.

\begin{figure}[t]
 \centering
 \includegraphics[angle=0,width=0.45\textwidth]{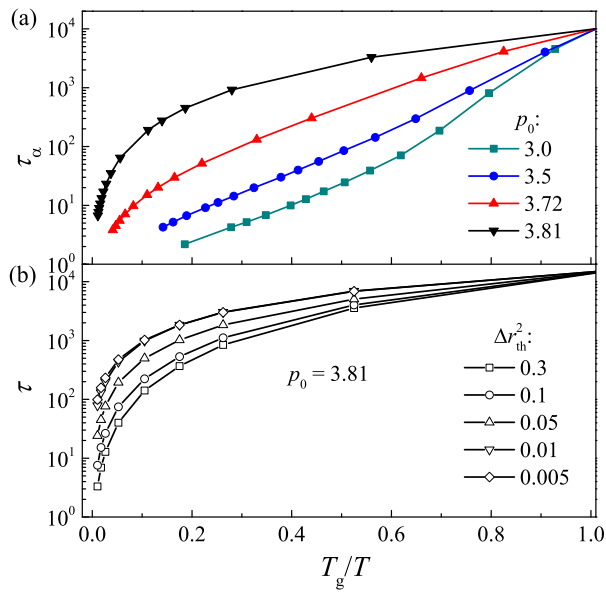}
 \caption{(a) Angell plot representation of the temperature dependence of the relaxation time, for different values of the target shape index $p_0$. $T_g$ is the glass transition temperature at which the relaxation time is $10^4$. (b) The Angell plot with the relaxation time defined as that where the mean square displacement attains a threshold value, as specified in the legend. The relaxation times are scaled so that they equal $\tau_\alpha$ at $T_g$.
\label{fig:Angell}
}
\end{figure}
The high-$p_0$ regime where the anomalous diffusive behaviour occurs, is also that where the relaxation time exhibits a sub-Arrhenius temperature dependence~\cite{Manning_EPL}, as we show in Fig.~\ref{fig:Angell}(a).
Our results obtained in an extended $p_0$ range, however, demonstrate that a traditional super-Arrhenius behaviour occurs at low $p_0$. 
A similar crossover is found defining the relaxation time from the decay of correlation function of the area and of the perimeter of the cells, as discussed in Appendix~\ref{sec:shape}.
A single parameter, $p_0$, thus controls the fragility and allows to transit from a sub- to a super-Arrhenius behaviour. 
We are not aware of other models with a similar crossover.

We now clarify that the anomalous sub-diffusive regime and the unusual sub-Arrhenius behaviour are strongly tied.
To this end, we define the relaxation time as that at which the mean square displacement reaches a threshold, $\langle \Delta r^2(\tau_{\Delta r^2_{\rm th}}) \rangle = \Delta r^2_{\rm th}$. 
Figures~\ref{fig:dyn}(a) and ~\ref{fig:dyn}(b) show that the mean square displacement at the relaxation time (circles) is $\Delta r^2_{\tau_\alpha} \simeq 0.1$, regardless of the temperature and of $p_0$. Hence, $\tau_{\Delta r^2_{\rm th}=0.1} \simeq \tau_\alpha$. 
The relaxation time $\tau_{\Delta r^2_{\rm th}}$ decreases with the threshold $\Delta r^2_{\rm th}$, becoming increasingly more influenced by the sub-diffusive regime rather than by the subsequent caging regime.
When this occurs, the sub-Arrhenius behaviour becomes more apparent, as we illustrate in Fig.~\ref{fig:Angell}(b).
We then conclude that the sub-diffusive behaviour induces the sub-Arrhenius one, while the caging regime contrasts it. 

\section{Anomalous glassy dynamics: physical origin}
\subsection{Particle trajectories}
The above results demonstrate that cell tissues, for large values of the target shape index $p_0$, have a distinctive relaxation dynamics, which is quite different from that of conventional glassy systems. Why is this so? 
To begin addressing this question, we have repeated the above investigations via overdamped simulations, for selected low-temperature values, and show the results as open circles in Figs.~\ref{fig:dyn}(a)-(d). 
These simulations reproduce the anomalous sub-diffusive regime, demonstrating that this has not an inertial origin. 
Besides, we have also investigated the relaxation dynamics using cage-relative quantities~\cite{Shiba, Keim_MW, Weeks_longwave, OurPNAS}, an approach which allows filtering out the effect of long-wavelength fluctuations.
These fluctuations might indeed be relevant in two spatial dimensions~\cite{MWPrl, OurPNAS, Keim_MW, Weeks_longwave, Shiba}, influencing both the mean square 
displacement and 
the relaxation time.
We illustrate in Appendix~\ref{sec:cr} that there are no considerable differences between the standard and the cage-relative relaxation dynamics; the anomalous sub-diffusive behaviour, therefore, is not the vestige of the vibrational dynamics of the system.

\begin{figure}[!t]
 \centering
 \includegraphics[angle=0,width=0.5\textwidth]{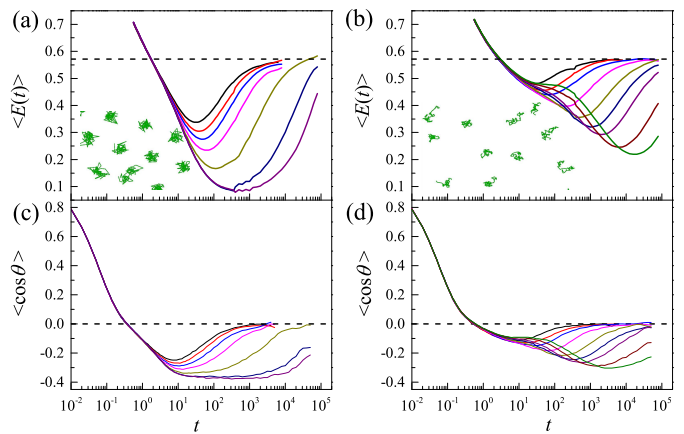}
 \caption{Time dependence of the averaged eccentricity $\langle E(t) \rangle$ and of $\langle \rm{cos} \theta \rangle$ (see text) at $p_0=3.0$, (a) and (c), and at $p_0=3.81$, (b) and (d).
The insets in (a) and (b) illustrate particle trajectories at the relaxation time. The horizontal dashed lines in (a) and (b) indicate the Brownian limit, $\langle E(t) \rangle =4/7$.
\label{fig:walker}
}
\end{figure}
To unveil the microscopic origin of the sub-diffusive behaviour, we then focus on the particle trajectories in the supercooled regime.
Example trajectories, evaluated at the relaxation time, are in Figs.~\ref{fig:walker}(a) for $p_0=3.0$, and \ref{fig:walker}(b) for $p_0=3.81$. 
We find that at small $p_0$, the trajectories have a round shape reflecting a caging regime, while conversely at large $p_0$ they are unusually stretched.
To quantify this observation, we describe a trajectory as a sequence of $n_v = 50$ points equally spaced in time. 
The eigenvectors of the gyration tensor of this set of points fix the spatial directions along which the fluctuations of the trajectory, as estimated by the squared eigenvalues $\lambda_1^2 \geq \lambda_2^2$, are maximal and minimal. 
This allows associating to each trajectory an eccentricity, $E(t)=\left(\lambda_{1}^{2}(t)-\lambda_{2}^{2}(t)\right)^{2}/\left(\lambda_{1}^{2}(t)+\lambda_{2}^{2}(t)\right)^{2}$~\cite{Gaspari1987,Matthias2012}. 
Radially symmetric trajectories have $E = 0$, straight lines $E=1$, while Brownian trajectories have $E=4/7$, in two dimensions~\cite{Gaspari1987}.
Consistently, the sample averaged eccentricity attains large values in the ballistic or super-diffusive regimes, reaches the Brownian limit at long times, and it is suppressed in the caging regime, as illustrated in Figs.~\ref{fig:walker}(a) and \ref{fig:walker}(b).
At $p_0 = 3.81$, an intermediate regime occurs between the ballistic and the caging one, where the eccentricity has a plateau.
The time dependence of the eccentricity, henceforth, closely resembles that of the log-slope, as apparent comparing Figs.~\ref{fig:walker}(a) and \ref{fig:walker}(b) with Figs.~\ref{fig:dyn}(c) and \ref{fig:dyn}(d).
More importantly, the trajectories reveal that the sub-diffusive behaviour results from an anisotropic motion of cells.
These anisotropic motion does not correlate with the possible anisotropic shape of the cell, as we show in Appendix~\ref{sec:correlation}.

The stretched trajectories lead to a sub-diffusive dynamics, rather than to a super-diffusive dynamics as one might naively expected, due to the presence of anti-correlations in the motion of the cells.
To highlight these correlations, we investigate the angle $\theta$ between consecutive displacements ${\bf r}(t_0+t)-{\bf r}(t_0)$,${\bf r}(t_0+2t)-{\bf r}(t_0+t)$, over a time $t$.
Both at low- and at high-$p_0$ values, the time evolution of $\langle \cos \theta \rangle$ resembles that of $\Delta(t)$ and of $E(t)$, as shown in Figs.~\ref{fig:walker}(c) and \ref{fig:walker}(d).
In particular, at large $p_0$, we observe an intermediate regime in between the ballistic and the caging ones.
In this intermediate regime, $\langle \cos \theta \rangle \approx -0.1$ for a transient.
Since this small value of $\langle \cos \theta \rangle$ occurs when the trajectories are elongated, we understand that sub-diffusion emerges as cells are transiently only slightly constrained.

\begin{figure}[!t]
 \centering
 \includegraphics[angle=0,width=0.45\textwidth]{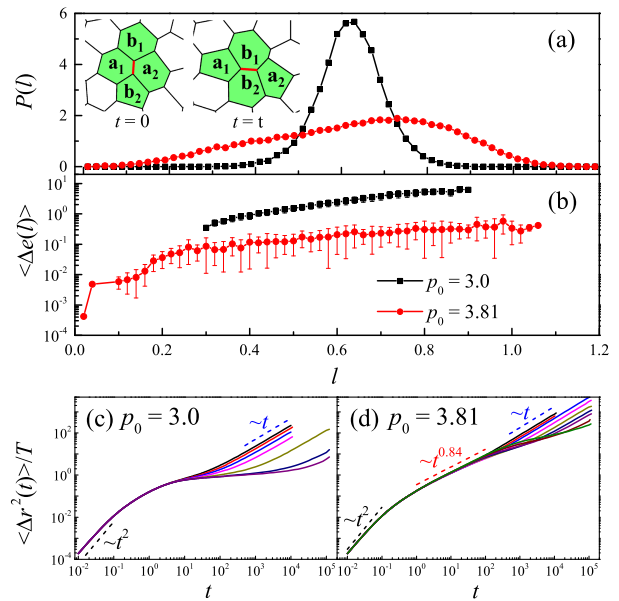}
 \caption{
 (a) Probability distribution of the cell edge length $l$ of energy minima configurations, and (b) dependence of the average energy barrier of T1 transitions on $l$, for $p_0=3.0$ (black squares) and $p_0=3.81$ (red circles). (a, inset): schematic of a T1 transition in which the edge connecting particles $\rm {a_1}$ and $\rm {a_2}$ disappears, and a novel edge connecting previously separated cells appears. (c) and (d) illustrate the time dependence of mean square displacement scaled by the temperature, respectively for $p_0=3.0$ and $p_0=3.81$. Colors indicate different temperature values, as in Fig.~\ref{fig:dyn}.
 \label{fig:edge}
}
\end{figure}

\subsection{T1 transitions}
The structural relaxation dynamics is strongly correlated with the topology of the free-energy landscape in glassy systems \cite{RevModPhys.83.587}.
Indeed, we now show the existence of phase space directions along which the system is essentially free to diffuse, for short distances, considering the energetic cost of relaxation events involving cell rearrangements, or T1 transitions~\cite{Rivier_1984, Staple2010, Bi_softMatter}, one of which is schematically illustrated in the inset of Fig.~\ref{fig:edge}(a).
In a T1 transition a cell-edge of length $l$ disappears, as the system overcomes an energy barrier $\Delta e(l)$ we expect to increase with $l$, as observed in the Vertex model~\cite{Bi_softMatter}.
We have investigated the edge-length distribution $P(l)$ and the dependence of the average energy barrier on $l$, which are illustrated Figs.~\ref{fig:edge}(a) and \ref{fig:edge}(b).
We detail the procedure used to evaluate these quantities is Appendix~\ref{sec:DE}.

At small $p_0$, $P(l)$ is Gaussian shaped, as observed in the Vertex model~\cite{Bi_softMatter}, and the average energy barrier increases with $l$.
At large $p_0$, $P(l)$ is broad and has almost a bi-modal shape, which is actually observed at even larger $p_0$ values not considered here~\cite{Ourwork_PRM}. 
In particular, on increasing $p_0$ small $l$-values become more probable. 
The energy cost of T1 transitions involving small edges, e.g. $l \lesssim 0.2$, is sensibly smaller than the energy cost of the other edges, as apparent in Fig.~\ref{fig:edge}(b).
Hence, the system is essentially free to diffuse along the specific phase space directions that trigger the T1 transitions involving these small edges.
To corroborate this picture we further consider that, since the free diffusion coefficient is proportional to $T$, the mean square displacement should scale as $T$ not only in the ballistic regime but also in the sub-diffusive one. 
We indeed observe in Figs.~\ref{fig:edge}(c) and \ref{fig:edge}(d) that, at low $p_0$, plots of $\left\langle\Delta r^{2}(t)\right\rangle/T$ only collapse in the ballistic regime, while conversely at high-$p_0$ they also collapse in the sub-diffusive one. 
The emerging scenario reminds the diffusion of a particle in a random media as described by the Lorentz gas models~\cite{Henk1982, Erwin2006, Franosch2010, Zeitz2017, Franosch2019}, where a particle is free until it hits randomly placed obstacles, and sub-diffusion occurs below the correlation length of the fractal cluster of free space.
\begin{figure}[t]
 \centering
 \includegraphics[angle=0,width=0.5\textwidth]{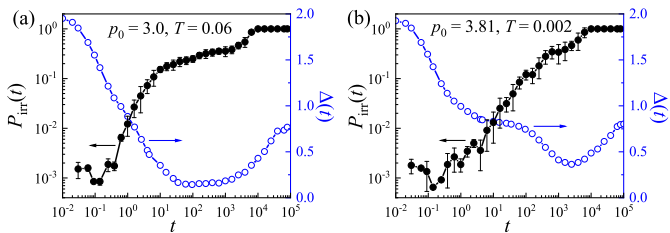}
 \caption{
 Time dependence of probability of irreversible consecutive T1 transitions (black) and of the log-slope of mean square displacement (blue) for (a) $p_0=3.0$ and $T=0.06$ and for (b) $p_0=3.81$ and $T=0.002$.
\label{fig:T1}
}
\end{figure}

To further support the deep connection between anomalous dynamics and T1 transitions, 
we consider the probability $P_{\rm irr}$ that two {\it consecutive} T1 transitions of the same particle are not one the reverse of the other; this occurs if the two transitions lead to a change in the Voronoi neighbours of the particle. 
We illustrate in Fig.~\ref{fig:T1} the dependence of $P_{\rm irr}$ on the time interval $t$ separating the two transitions. 
To avoid cluttering of data, we consider in (a) $p_0 = 3.0$ and $T =0.06$, and in (b) $p_0 = 3.81$ and $T = 0.002$, two state points having close relaxation time, and report results for other parameter values in Appendix~\ref{sec:T1}. 
In the figure, we also superimpose the log-slope $\Delta(t)$ of the mean square displacements.

For small $p_0$, $P_{\rm irr}$ quickly attains a high, almost constant plateau value, characterizing the caging regime.
$P_{\rm irr}$ then approaches $1$ as the system relaxes.
For large $p_0$, $P_{\rm irr}$ grows essentially as a power-law during the sub-diffusive transient. 
An inflexion, reminiscent of a plateau in the caging-regime follows the power-law growth and the final approach to $1$. 
Hence, the sub-diffusive regime is characterized by a scarcity of irreversible transition.

\section{Discussion}
Our study demonstrates that the relaxation dynamics of a model cell tissue qualitatively depends on the stiffness of the cells; 
while stiff cells exhibit a conventional glass-like relaxation dynamics, soft ones have an extended sub-diffusive transient and a sub-Arrhenius dependence on the relaxation time on the temperature. 
Consistently, dynamical heterogeneities grow on cooling for stiff cells, while they are almost temperature independence for soft cells, as we demonstrate in Appendix~\ref{sec:length}.
The qualitative changes in the relaxation dynamics originate from the emergence of phase space directions along which the system is essentially free to move, in soft cells, and establish an analogy between the energy landscape of cell tissues and the Lorentz model, on short length scales. 

We do not expect the sub-diffusive behavior we have discussed to be a universal feature of the dynamics of cell tissues.
Its occurrence, indeed, might be hidden by the super-diffusive contribution to the mean square displacement of the active forces.
To observe our finding one might suppress cell-motility, making the cell tissue dynamics thermal.
In order for thermal forces alone to be able to induce the relaxation of the system, it migth be also convenient to consider soft tissues, as those close to the epithelial-mesenchymal transition~\cite{EMT1, EMT2, EMT3}.

We remark, however that a sub-diffusive transient, $r^2 \propto t^\beta$, with $\beta$ constant over an extended period of time, has been observed in some experiments~\cite{Rogers2007, Manning2013, Nixon-Abell2016, Armiger2018, Fodor2018}.
Our results offer a possible explanation of these experimental findings because at short time the thermal contribution to the mean square displacement ($\propto t$), which is the one we have modeled, dominates over the active contribution ($\propto t^2$).



\begin{acknowledgments}
We acknowledge support from the Singapore Ministry of Education through the Academic Research Fund MOE2017-T2-1-066 (S), and are grateful to the National Supercomputing Centre (NSCC) of Singapore for providing computational resources. 
MP is supported by the H2020 program under the MSCA grant agreement No. 801370 and by the Secretary of Universities and Research of the
Government of Catalonia through Beatriu de Pin\'os program Grant No. BP 00088 (2018).
\end{acknowledgments}

\appendix
\section{Shape correlation functions\label{sec:shape}}
\begin{figure}[htb]
 \centering
 \includegraphics[angle=0,width=0.5\textwidth]{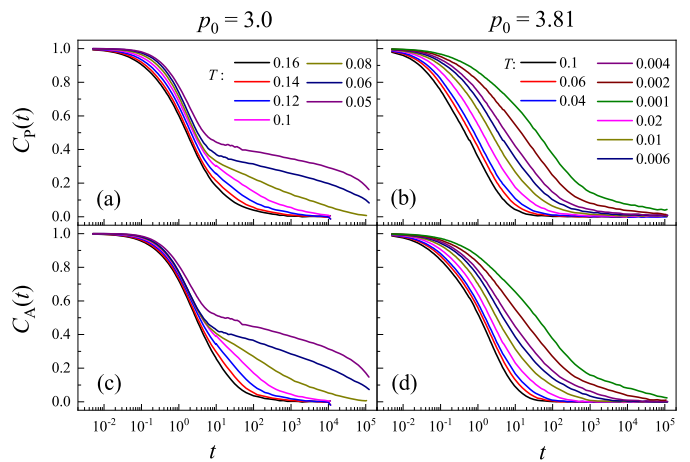}
\caption{Time dependence of perimeter (a) and of area (c) correlation functions at $p_0=3.0$. (b) and (d) are the time evolution of the same quantities at $p_0=3.81$.
\label{fig:Cshape}}
\end{figure}

To prove that the anomalous dynamics are associated to changes in the shapes of the cells, as those one might expect T1 transitions to induce, we investigate the perimeter and area correlation functions. 
The perimeter correlation function is defined as
\begin{equation}
C_{P}(t)=\frac{\sum_{i=1}^{N}\left[p_{i}(t)- \langle p_{i}\rangle \right]\left[p_{i}(0)-\langle p_{i}\rangle\right]}{\sum_{i=1}^{N}\left[p_{i} (0)-\langle p_{i}\rangle\right]^{2}},
\label{eq:Cp}
\end{equation}
where $p_{i}(t)$ is the non-dimensional perimeter of cell $i$ at time $t$ and $\langle p_{i}\rangle$ is the time average value, which is cell dependent due to the polydispersity of our system.
The area correlation function $C_{A}(t)$ is similarly defined.

We illustrate the time dependence of $C_{P}(t)$ and of $C_{A}(t)$ at $p_0=3.0$ and at $p_0=3.81$ in Fig.~\ref{fig:Cshape}. 
$C_{P}(t)$ and $C_{A}(t)$ demonstrate similar behavior. 
In particular, at $p_0=3.0$, both correlation functions exhibit a two-step decay, which is conversely not apparent at $p_0=3.81$. 
This $p_0$ dependence is consistent with that of the ISF (Fig. 1) and CR-ISF (Figs.~\ref{fig:crdyn}(e) and \ref{fig:crdyn}(f)).
We further extract from the shape correlation function the perimeter and the area relaxation time, $\tau_P$ and $\tau_A$, which satisfy $C_{P}(\tau_P)=C_{A}(\tau_A)=1/e$.
Both relaxation time have a super-Arrhenius temperature dependence at small $p_0$, and a sub-Arrhenius one at large $p_0$, as we show in Fig.~\ref{fig:Angellalltau}(b). 
The investigation of the relaxation dynamics via the shape-correlation function establishes a coupling between the geometrical properties of the cells and their displacement.
Considering that the shape of the cell changes as a consequence of T1 transitions, this result indirectly links anomalous dynamics and T1 transitions.

\section{Cage-relative dynamics\label{sec:cr}}

\begin{figure}[htb]
 \centering
 \includegraphics[angle=0,width=0.5\textwidth]{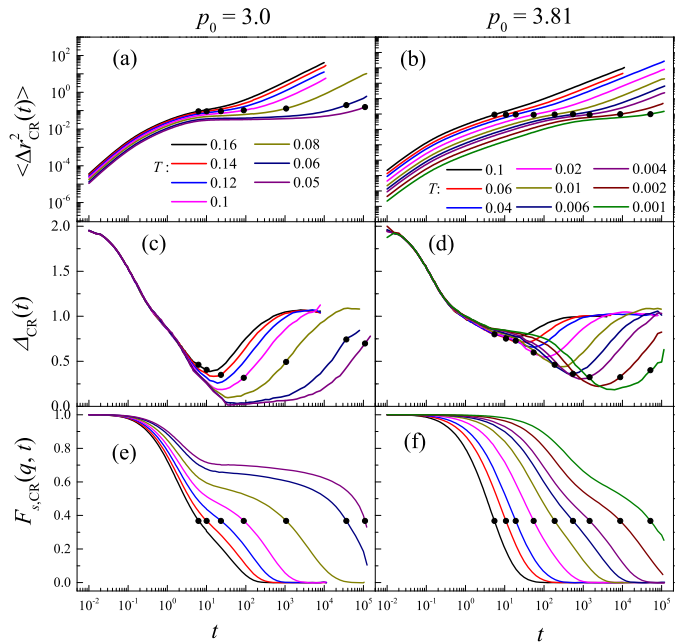}
\caption{Time dependence of the cage-relative mean square displacement (a), of its log-slope (c) and of the cage-relative self-intermediate scattering function (e) at $p_0=3.0$. (b), (d) and (f) show the time dependence of the same quantities at $p_0=3.81$. The full circles in all panels mark the cage-relative relaxation time  $\tau_\alpha^{\rm CR}$, which is the time at which the cage-relative self-intermediate scattering function reaches $1/e$.
\label{fig:crdyn}}
\end{figure}

\begin{figure}[htb]
 \centering
 \includegraphics[angle=0,width=0.5\textwidth]{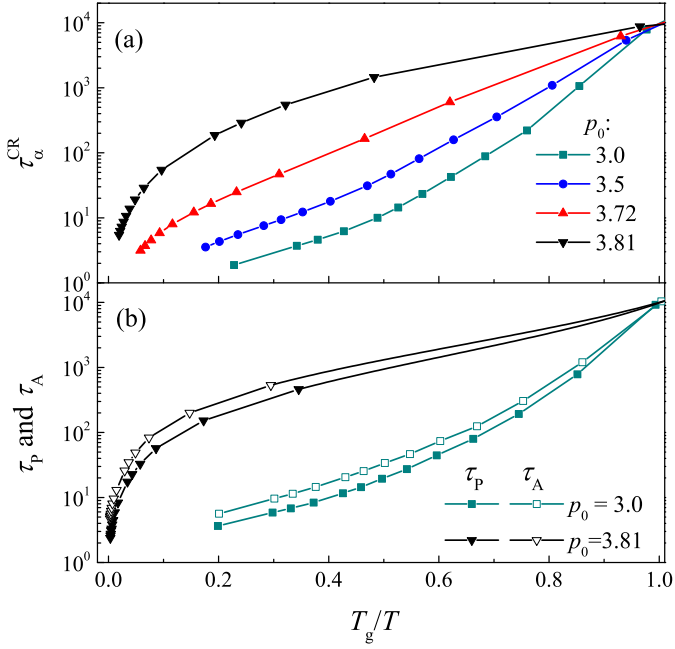}
\caption{Angell plots, as obtained using different definitions of the relaxation time. In (a), $\tau_\alpha^{\rm CR}$ is the cage-relative relaxation time. In (b), $\tau_{\rm P}$ and $\tau_{\rm A}$ are the perimeter and the area relaxation time.
\label{fig:Angellalltau}}
\end{figure}

We have illustrated in Fig. 1 the mean square displacements (MSD), its log-slope, and the self-intermediate scattering function (ISF). 
We have additionally investigated the time dependence of these quantities using cage-relative (CR) measures.
The CR measures differ from the standard ones in that the CR displacement $\Delta \mathbf{r}_{i}^{\rm CR}(t) =
\Delta \mathbf{r}_i(t)
 - 1/N_{i}\sum_{j=1}^{N_{i}} \Delta \mathbf{r}_j(t)$, where the sum is over the $N_{i}$ neighbors particle $i$ has at time $0$, replaces the displacement $\Delta \mathbf{r}_i(t) = \mathbf{r}_i(t)-\mathbf{r}_i(0)$.
Particles moving coherently with their immediate neighbours have a large displacement, but a small CR displacement. 
Hence, CR measures filter out the effect of coherent displacements, and particularly the effect of long-wavelength fluctuations, which could affect the relaxation dynamics of two-dimensional systems~\cite{Shiba, Weeks_longwave, Keim_MW, OurPNAS}.

Figure~\ref{fig:crdyn} shows that, for small $p_0$, the CR one reveals a typical glassy behaviour, including an extended plateau in CR-MSD and a two-step decay in CR-ISF at low temperatures, as the standard measure.
Similarly, the anomalous sub-diffusive behaviour found at large $p_0$ persists when the relaxation dynamics is investigated using CR measures. 
Indeed, a region of anomalous diffusion is clearly observed in CR-MSD (Fig.~\ref{fig:crdyn}(b)) and in its log-slope (Fig.~\ref{fig:crdyn}(d)).
This anomalous behaviour, and that observed in the standard quantities in Fig. 1  at the same $p_0$ value, occur on the same time scale.

We further define the CR relaxation time $\tau_\alpha^{\rm CR}$ as the time at which CR-ISF reaches $1/e$, and illustrate the resulting Angell plot in Fig.~\ref{fig:Angellalltau}(a).
On increasing $p_0$, we observe a crossover from a super- to a sub-Arrhenius behaviour, as found in Fig. 2(a) using the standard measure.

Overall, the investigation of the relaxation dynamics using CR quantities excludes the possibility that the observed anomalous behaviour occurring at large $p_0$ could originate from the emergence of collective particle displacements, like those induced by long-wavelength fluctuations.

\section{Absence of correlation between shape and  displacement of a cell\label{sec:correlation}}
\begin{figure}[!tb]
 \centering
 \includegraphics[angle=0,width=0.5\textwidth]{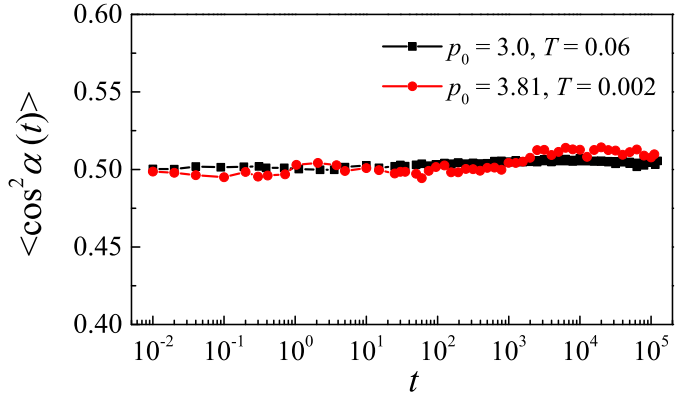}
\caption{Time dependence of $\langle \cos^2 \alpha(t) \rangle$ for $p_0=3.0$ and $T=0.06$ (black squares) and for $p_0=3.81$ and $T=0.002$ (red circles).
\label{fig:corr}}
\end{figure}

Cells in tissue, being deformable objects, may acquire elongated shapes.
A cell's displacement could, therefore, correlate with its shape, e.g. in the anomalous diffusive regime at high $p_0$.
To investigate this possibility, we first assume the eigenvector associated with the largest eigenvalue of the covariance matrix of the vertices of cell $i$ to identify its principal axis, $\mathbf{v}_i$.
Next, we consider how the normalized cell displacement $\Delta \mathbf{r}_i (t)/|\Delta \mathbf{r}_i (t)|$ at time $t$ correlates with the principal axis at time $0$, studying $\cos \alpha_i(t) = \Delta \mathbf{r}_i (t)/|\Delta \mathbf{r}_i (t)| \cdot \mathbf{v_i}(0)$.
Since the cell's principal axis is defined up to an angle $\pi$,
$\langle \cos \alpha(t) \rangle = 0$.
We, therefore, focus on $\langle \cos^2 \alpha(t) \rangle$, which equal $1/2$ in the absence of correlations.
In Fig.~\ref{fig:corr}, we show that $\langle \cos^2 \alpha(t) \rangle$ does equal $1/2$, regardless of the $p_0$ value and of the time. 
Analogous results are obtained at different temperatures.
Accordingly, the shape of a cell at a given time does not correlate with its subsequent displacement. 
This result is consistent with our finding, according to which in the anomalous region cells move along direction inducing T1 transition associated with their short edges.

\section{T1 energy barrier\label{sec:DE}}
\begin{figure}[htb]
 \centering
 \includegraphics[angle=0,width=0.5\textwidth]{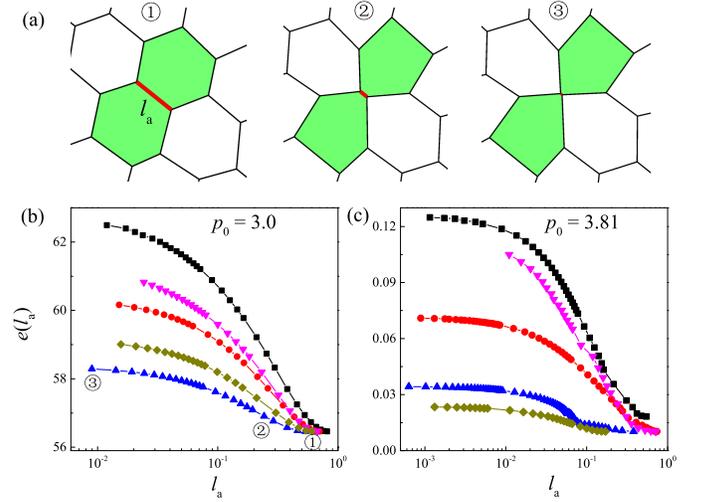}
\caption{(a) Illustration of tuning the edge length $l_{{\rm a}}$ between two selected neighboring cells (green) by separating them gradually so as to induce a T1 transition. (b) and (c) the edge length $l_{{\rm a}}$ dependence of the minimised total energy $e(l_{{\rm a}})$ at $p_0=3.0$ and $p_0=3.81$, respectively. Different colors are for different selected neighboring cell couples. The edge length $l_{{\rm a}}$ and the corresponding energy $e(l_{{\rm a}})$ for the snapshots shown in (a) are indicated in (b).
\label{fig:Eb}}
\end{figure}

We investigate the energy barrier for T1 transition to occur focusing on systems with $N = 100$ cells quenched to their inherent state via the conjugate-gradient algorithm.
In these systems, we randomly select two neighbouring cells and indicate with $l_{{\rm a}}$ the length of the Voronoi edge separating them.
Then, we gradually increase the separation of the two cells, moving them by small steps along the direction connecting their centres. 
We fix the step size to $0.1$, $0.01$, and $0.001$ when the distance $dr$ between the cell centers is $dr>0.2$, $0.2>dr>0.1$, and $0.1>dr$, respectively.
After each step, we minimize the energy of the tissue using the conjugate-gradient method, keeping fixed the positions of the selected cells.
As the distance between the centers of selected cells increases, the length $l_{\rm a}$ of the Voronoi edge separating them decreases, and the energy of the system increases, as visualized in Fig.~\ref{fig:Eb}(a).
As the length scale increases, the energy of the tissue grows, as illustrated for a few selected cell couples in Fig.~\ref{fig:Eb}(b) for $p_0 = 3.0$, and in Fig.~\ref{fig:Eb}(c) for $p_0=3.81$. 
The energy suddenly drops as the T1 transition separating the selected particles occurs, as $l_{{\rm a}}$ approaches $0$.
The overall change in energy defines the energy barrier $\Delta e(l) = e(l_{{\rm a}}\rightarrow0) - e(l)$. 
Figure 4(b) illustrates $\langle \Delta e(l) \rangle$ as a function of the initial edge length $l$. The data are obtained randomly by triggering 200 random T1 transitions, from 24 independent configurations.

We note here that in a few instances we have observed drops in the dependence of the energy versus $l_{{\rm a}}$ due to T1 transitions which do not involve the displaced particles. Regardless, we operatively define $\Delta e(l)$ as the difference between the energy of the system as the separating particles undergo a T1 transition and the initial one.

\section{T1 correlations\label{sec:T1}}
In Fig.~\ref{fig:T1}, we illustrate the time dependence of the probability to find irreversible consecutive T1 transitions, $P_{\rm irr}(t)$, at different temperatures for $p_0=3.0$ (panel (a)) and for $p_0=3.81$ (panel (b)).
Figure 5 shows that data for $p_0=3.0$ and $T=0.06$ and for $p_0=3.81$ and $T=0.002$ are qualitatively different, and that $P_{\rm irr}(t)$ correlates with the MSD.

Here, we notice that the temperature dependence of $P_{\rm irr}(t)$ is qualitatively the same, for different $p_0$ values. 
At higher temperature, it becomes increasingly more probable for two consecutive transitions separated by a small time interval $t$ not to be one the reverse of the other. Furthermore, as the temperature increases the plateau that $P_{\rm irr}(t)$ attains at long-time during the caging regime, reduces in extension and increases in value approaching $1$.
\begin{figure}[tb]
 \centering
 \includegraphics[angle=0,width=0.5\textwidth]{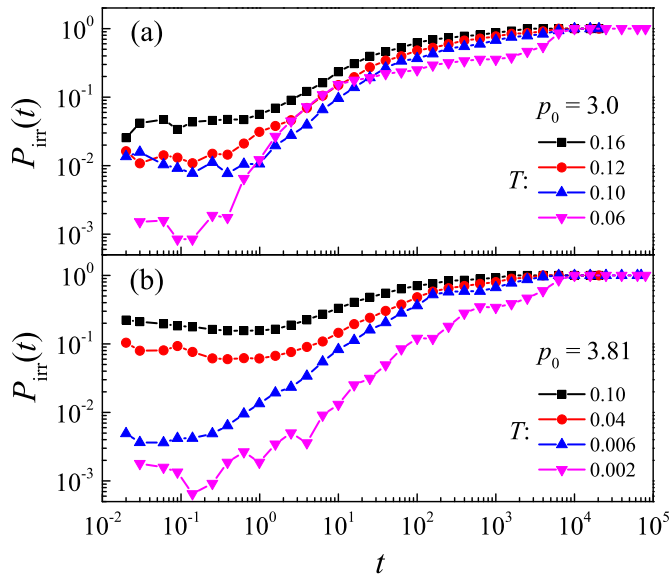}
\caption{Time dependence of the probability that two consecutive T1 transition of a same particle are not one the reverse of the other, for several selected values of temperatures at (a) $p_0=3.0$ and at (b) $p_0=3.81$.
\label{fig:T1}}
\end{figure}

\section{Dynamical length scales\label{sec:length}}

\begin{figure}[!h]
 \centering
 \includegraphics[angle=0,width=0.5\textwidth]{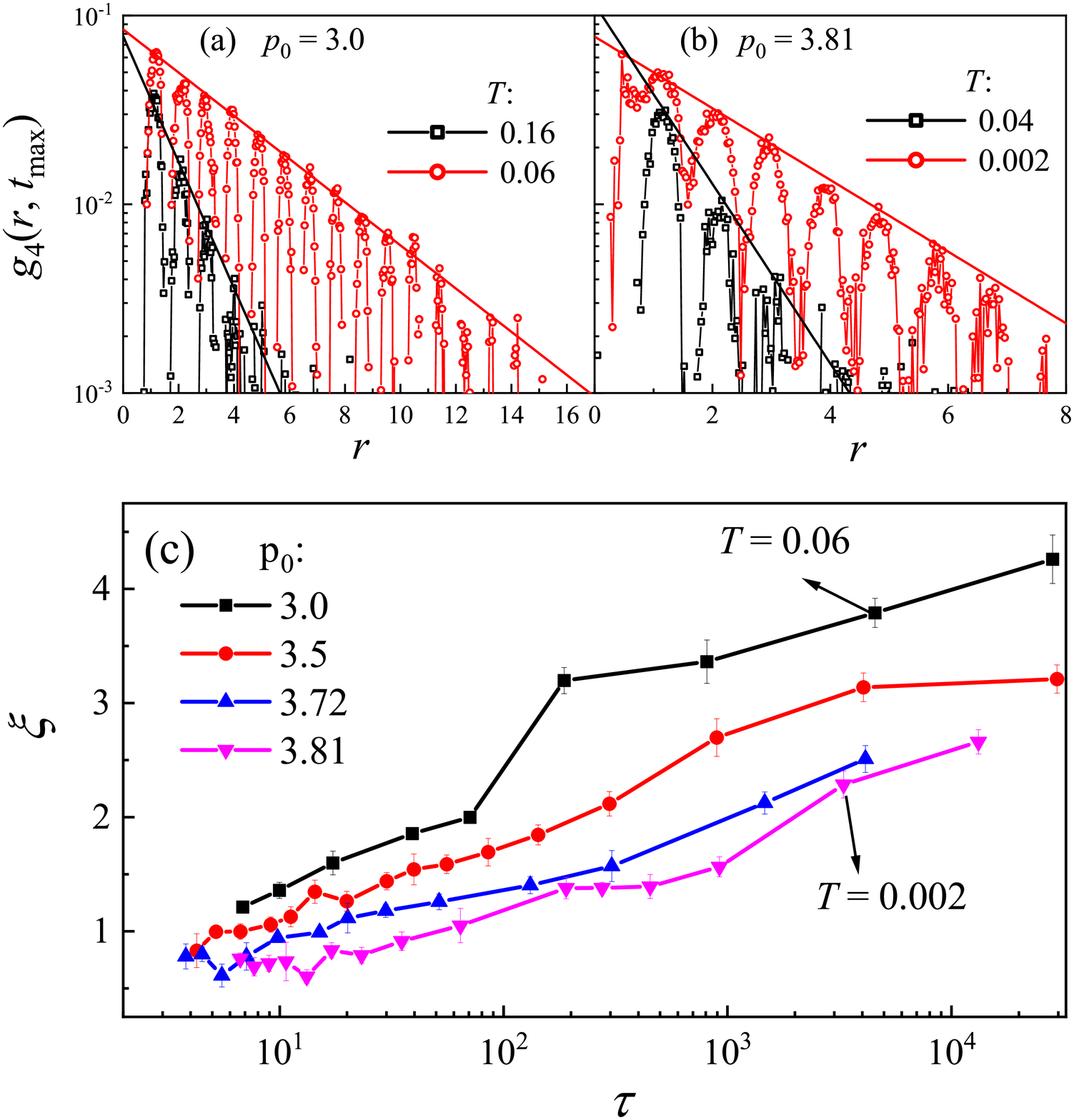}
\caption{Spatial-temporal correlation functions at time $t_{\rm max}$ for (a) $p_0=3.0$ and for (b) $p_0=3.81$.
$t_{\rm max}$ is the time at which the corresponding four-point susceptibility reaches the maximum. 
The solid lines are exponential fits. 
(c) illustrates the dependence of the dynamical correlation length on the relaxation time for different $p_0$ values.
\label{fig:length}}
\end{figure}

The existence of a standard and of an anomalous glassy dynamics, respectively at small at a high $p_0$ values, suggests that the spatial temporal correlation between the particle displacement may likewise be strongly $p_0$ dependent.

To investigate this issue, we focus on the decay of the spatial-temporal correlation function~\cite{Pastore,Kites}:
\begin{equation}
g_{4}(r_{ij},t)=\langle\omega_{i}(t)\omega_{j}(t)\rangle -
\langle\omega_{i}(t)\rangle \langle\omega_{j}(t)\rangle.
\label{eq:g4}
\end{equation}
Here $r_{ij}=|\mathbf{r}_{i}(0)-\mathbf{r}_{j}(0)|$ and $\omega_{i}(t)=1(0)$ if $|\mathbf{r}_{i}(t)-\mathbf{r}_{i}(0)|\leq$ ($>$) $l_{*}$. 
We fix $l_{*} = 0.64$, the value at which the peak height of the corresponding four-point susceptibility $\chi_{4}(t)=1/N\sum_{i,j}g_{4}(r_{ij},t)$ is maximal, and fix the time $t=t_{\rm max}$ at which the corresponding $\chi_{4}(t)$ attains the maximum.
From the exponential decay of $g_{4}(r,t)$, which is illustrated in 
Figs.~\ref{fig:length}(a) and \ref{fig:length}(b) for selected $p_0$ and temperature values, we then extract the dynamical length scale $\xi$.

In Fig.~\ref{fig:length}(c), we illustrate the dependence of the length scale $\xi$ on the relaxation time, for different values of $p_0$. 
$\xi$ increases as the dynamics slow down.
At a given relaxation time, we observe a systematic reduction of $\xi$ on increasing $p_0$. 
This indicates that dynamic heterogeneities decrease as the softness of the particles increases, in line with the absence of a proper glassy behavior for these particles.


%
\end{document}